\documentclass[prl,amsfonts,english,showpacs,reprint,floatfix]{revtex4-1}
\usepackage[T1]{fontenc}
\usepackage[latin1]{inputenc}
\usepackage{amsmath}
\usepackage{graphicx}
\usepackage{amssymb}
\usepackage{verbatim}

\usepackage{soul}
\makeatletter
\usepackage{babel}
\usepackage[normalem]{ulem} \makeatother
\newcommand{\eref}[1]{Eq.\ (\ref{#1})}
\newcommand{\Eref}[1]{Equation (\ref{#1})}
\newcommand{\fref}[1]{Fig.\ \ref{#1}}

\begin{document}

\title{Phase diagram and density large deviations of a nonconserving ABC model}

\author{O. Cohen and D. Mukamel}

\address{Department of Physics of Complex Systems, Weizmann Institute of Science,
76100 Rehovot, Israel}

\date{\today}

\begin{abstract}
The effect of particle-nonconserving processes on the steady state
of driven diffusive systems is studied within the context of a
generalized ABC model. It is shown that in the limit of slow
nonconserving processes, the large deviation function of the overall
particle density can be computed by making use of the steady state
density profile of the conserving model. In this limit one can
define a chemical potential and identify first order transitions via
Maxwell's construction, similarly to what is done in equilibrium
systems. This method may be applied to other driven models subjected
to slow nonconserving dynamics.
\end{abstract}
\maketitle

Driven diffusive systems have been at the focus of extensive
theoretical and experimental studies in recent years. Many studies
have been devoted to systems of particles evolving under local
biased exchange processes in the bulk which may or may not be
coupled to nonconserving reservoirs at the boundaries. Examples
include the asymmetric exclusion process (ASEP), the zero-range
process (ZRP) and many others \cite{Mukamel2000,Schutz2001, Evans2005, Schadschneider2010}.
Bulk-nonconserving processes are rather common in many physical systems such as
molecular motors, traffic flow problems with road intersections, chemical reactions in an open
environment and others (see for example \cite{Schadschneider2010}).
The corresponding driven models of such systems are often less
tractable analytically
\cite{Evans2002,Willmann2002,Parmeggiani2003,Evans2003,Popkov2003,Levine2004,Sasa2006,Bodineau2010}.
Their steady-state properties are usually very different from
those of the corresponding conserving models.

In this Letter we study the effect of bulk nonconserving processes
 by considering the limit where they occur on a much longer time scale than that of
 the conserving dynamics.
 This separation of time scales allows the system to relax to its
conserving steady state with a fixed number of particles in between
nonconserving dynamical moves. As a result, the steady state of the
system can be expressed in terms of an appropriate `ensemble' of the
conserving steady states which may be regarded as a kind of
nonequilibrium grand-canonical ensemble. One may thus use the steady
state properties of the conserving system to analyze the
corresponding properties of the nonconserving one.

This approach is demonstrated on the ABC model \cite{Evans1998,Evans1998b}. This is a one
dimensional three species driven exclusion model which exhibits a
phase separation transition, and which has previously been
generalized to include particle-nonconserving processes. By considering the limit
of slow nonconserving dynamics and following the approach described above,
we compute explicitly the
large deviation function (LDF) of the overall particle density,
despite the fact that the LDF of the density profile is not known.
The LDF yields a definition for a `chemical potential' which unlike the equilibrium
one, depends on the details of the nonconserving dynamical process.
Based on this derivation, we draw the exact phase diagram of the model.
The first
order transition line exhibited by the model may be computed via
Maxwell's construction, even though detailed balance is not obeyed.
As discussed at the end of the Letter, the method presented below
is readily applicable to other driven models that are coupled
slowly to an external reservoir.

The ABC model is defined on a one-dimensional
periodic lattice of length $L$, where each site is occupied by one of the three species of particles,  labeled $A,\, B$ and $C$. The model evolves by random sequential updates whereby particles on neighbouring sites are exchanged with the following rates,
\begin{equation}
AB\overset{q}{\underset{1}{\rightleftarrows}}BA\,\qquad\,
BC\overset{q}{\underset{1}{\rightleftarrows}}CB\,\qquad\,
CA\overset{q}{\underset{1}{\rightleftarrows}}AC.\label{eq:ABCdynamics}
\end{equation}
For $q=1$, the model relaxes to an equilibrium steady state where the particles are homogeneously
 distributed.  For any finite value of $q\neq 1$, the model exhibits phase separation into three domains
 in the limit of $L\to \infty$.
 Generically,
the model does not obey detailed balance and it relaxes to a nonequilibrium steady-state.
A special feature of the ABC model is that in the case where the number of particles of the
three species are equal, $N_A=N_B=N_C$,
the dynamics obeys detailed balance with respect to an effective Hamiltonian with long-range interactions.

The model is often studied in the limit of weak asymmetry where
$q$ approaches $1$ in the thermodynamic limit as $q=\exp(-\beta/L)$ \cite{Clincy2003}.
This model exhibits a phase transition at some value of $\beta$ between a homogenous phase
and an ordered phase with
three macroscopic domains, each predominantly occupied by one of the species.
The transition point and its order depend on the values of $N_A,N_B$ and $N_C$ \cite{Clincy2003,Cohen2011a}.
In the equal-densities case the parameter $\beta$ plays the role of the inverse temperature.
This phase transition has also been studied by considering various generalization of the ABC model
such as interval boundary conditions \cite{Ayyer2009}, species-dependent $q$ \cite{Barton2011} and  particle-nonconserving dynamics \cite{Barton2010,Lederhendler2010a,Lederhendler2010b}.

In the following we analyze the generalized weakly-asymmetric ABC model on a ring with particle-nonconserving dynamics \cite{Lederhendler2010a,Lederhendler2010b}.
In this model sites can be occupied by inert vacancies, denoted by $0$, whose dynamics is defined as
\begin{equation}
A0\overset{1}{\underset{1}{\rightleftarrows}}0A,
\qquad B0\overset{1}{\underset{1}{\rightleftarrows}}0B,
\qquad C0\overset{1}{\underset{1}{\rightleftarrows}}0C.\label{eq:vacancyexchange}
\end{equation}
The total number of particles, $N=N_A+N_B+N_C\leq L$,
fluctuates through evaporation and deposition of triplets of neighbouring particles given by
\begin{equation}
ABC\overset{p \, e^{-3\beta \mu}}{\underset{p}{\rightleftarrows}}000.\label{eq:addremoverates}
\end{equation}
Here $p$ is the overall rate of the nonconserving dynamics and $\mu$ is a parameter
which as shown below, can be regarded as a chemical potential.
This type of nonconserving process has been chosen because it
 maintains detailed balance with respect to an effective Hamiltonian if initially the densities
 are equal, $N_A=N_B=N_C$.

In the equal-densities case, the equilibrium canonical and grand-canonical ensembles of this model
correspond respectively to %two models:
a {\it conserving model}, defined by rules (\ref{eq:ABCdynamics})
and (\ref{eq:vacancyexchange}) and a {\it nonconserving model}, which includes also rule (\ref{eq:addremoverates}).
It has been found that for this case both models exhibit the same second order transition line, which turns into a first order line at a tricritical point only in the nonconserving model \cite{Lederhendler2010a,Lederhendler2010b}.
Such inequivalence between the phase diagrams of the two ensembles is often observed in long-range interacting systems (for recent reviews see \cite{Dauxois2009,Dauxois2010}).

In this Letter we study the phase diagram of the generalized weakly-asymmetric ABC model
with nonequal densities. In this case no free energy exists. This study can be accomplished by
analyzing the steady state of its hydrodynamic equations, given by
\begin{align}
\label{eq:meanfieldA}
\partial_{t} \rho_\alpha = \, &  \beta L^{-2}\partial_x\left[\rho_\alpha\left(\rho_{\alpha+1}-\rho_{\alpha+2}\right)\right]+ L^{-2}\partial_x^2\rho_\alpha \nonumber \\
& +  p\left(\rho_0^3-e^{-3\beta\mu}\rho_A\rho_B\rho_C\right).
\end{align}
Here $\rho_\alpha\left(x\right)$
is the coarse-grained density profile whose index $\alpha$ denotes the species and runs cyclicly over $A\,,B$ and $C$. We denote the average density of each species by
 $r_\alpha=N_\alpha/L=\int_0^1dx \rho_\alpha \left(x\right)$ and the overall density by
$r=r_A+r_B+r_C$.
In the original ABC model ($r=1,p=0$), \eref{eq:meanfieldA} has been shown to be exact in the limit of
 $L\to\infty$ for equal densities \cite{Clincy2003,Ayyer2009},
and has been argued to remain so even for arbitrary average densities \cite{Ayyer2009}.

The conserving steady state ($r\leq 1,p=0$) of \eref{eq:meanfieldA}, denoted by $\rho^\star_\alpha\left(x,r\right)$,
 can be readily extracted from the known steady state of the original ABC model,  $\rho^\star_\alpha(x,1)$, via a scaling transformation discussed below.
In the nonconserving model, we are able to derive the steady-state of
\eref{eq:meanfieldA} in the limit of slow nonconserving dynamics,
\begin{equation}
\label{eq:slowp}
p\sim L^{-\gamma},\qquad \gamma > 2,
\end{equation}
where the $p$-dependent term becomes subdominant.
Dynamically this limit implies that on time scales of order $L^2$ the nonconserving model relaxes to the
conserving steady state with a fixed overall particle density, $r$, whereas
on longer time scales of order $L^{\gamma}$, $r$ fluctuates around its steady-state value.

 This separation of time scales suggests that \eref{eq:meanfieldA} remains valid within the limit considered in \eref{eq:slowp} and that the steady state measure of the nonconserving model can be written in the limit of large $L$ as
\begin{equation}
\label{eq:adiabatic}
P_{nc}\left( \boldsymbol \zeta,N\right) \simeq P_c\left( \boldsymbol \zeta \, ; N \right) P\left(N\right).
\end{equation}
Here ${\boldsymbol \zeta}=\{\zeta_i\}$ denotes a microstate of the model
with $\zeta_i=A,B,C$ for $i\in[1,L]$ and
$P_c\left( \boldsymbol \zeta \, ; N \right)$ is the conserving steady-state measure.
Although $P_c\left( \boldsymbol \zeta \, ; N \right)$ is not known, the knowledge
of its extremizing profile in the hydrodynamic limit, $\rho^\star_\alpha\left(x,r\right)$, is sufficient for deriving the probability density of $N$ in the nonconserving model, $P(N)$.

We now derive $P(N)$ by writing its master equation, which evolves on the slow time scale of the bath by
\begin{equation}
\label{eq:Master}
\partial_{t}P(N)=\sum_{N'=N\pm3}  P(N') Q(N'\to N)-P(N)Q(N \to N') .
\end{equation}
The average transition rates, $Q$, are given by
\begin{equation}
\label{eq:n_rates}
Q\left(N \to N'\right) =
\sum_{\boldsymbol \zeta} P_c \left(\boldsymbol \zeta \, ; N \right) \, \sum_{\boldsymbol \zeta',{N'}}
W\left( \boldsymbol \zeta,N | \boldsymbol \zeta',{N'}\right),
\end{equation}
where $N'=N\pm 3$ and $W$ is the transition rate between two microstates.
The latter is computed by counting the
number of $ABC$ and $000$ triplets in each microstate, denoted by $ n_{ABC} \left( \boldsymbol \zeta \right)$ and $n_{000} \left( \boldsymbol \zeta \right)$, respectively.
In the limit of $L\to\infty$ the evaporation rate
 can be computed using a saddle-point approximation, yielding
\begin{align}
\label{eq:n_rates1}
Q\left(N \to N-3\right) = p\, e^{-3\beta\mu} \sum_{\boldsymbol \zeta}  P_c \left(\boldsymbol \zeta \, ; N \right) n_{ABC} \left( \boldsymbol \zeta \right)  \nonumber \\
   \simeq  p \, e^{-3\beta\mu} L\int_0^1 \!\! dx\rho_A^\star\left(x,r\right)\rho_B^\star\left(x,r\right)\rho_C^\star\left(x,r\right).
\end{align}
The deposition rate is obtained similarly as
\begin{equation}
\label{eq:n_rates2}
Q\left(N \to N+3\right) \simeq p\,L\int_0^1 \!\! dx\left[\rho_0^\star\left(x,r\right)\right]^3,
\end{equation}
%Here $\rho_\alpha^\star\left(x,r\right)$ is the conserving steady-state profile with fixed density, %$r=N/L$.
which can be simplified by noting that the inert vacancies have a flat steady-state profile, $\rho_0^\star\left(x,r\right)=1-r$.

\begin{figure}[t]
\noindent
\begin{centering}\includegraphics[scale=0.46]{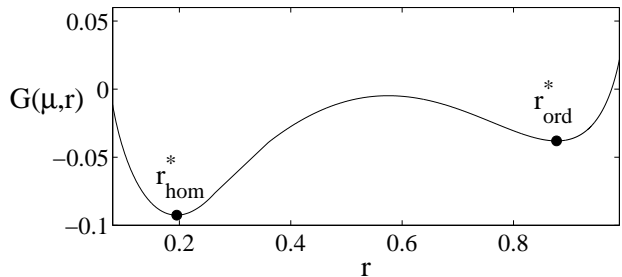}\par\end{centering}
\caption{ The large deviation function of the overall density, $r$, for $\beta=50$, $r_A=r_B=r/3-0.025$ and $\mu=-0.053$. For these parameters the LDF has two local minima, at $r^{\star}_{\text{hom}}$ and $r^{\star}_{\text{ord}}$, corresponding to the
homogenous and ordered phases, respectively.
The nonconserving model undergoes a first order phase transition when
$G(\mu,r^{\star}_{\text{hom}})=G(\mu,r^{\star}_{\text{ord}})$.
 %The first order transition point is defined by %$\mathcal{F}_\mu(r^{\star}_{\text{hom}})=\mathcal{F}_\mu(r^{\star}_{\text{ord}})$.
\label{fig:Fmu}}
\end{figure}

\Eref{eq:Master} corresponds
 to a one-dimensional random walk in $N$ in the presence of a local potential whose steady state is
\begin{equation}
\label{eq:PN}
P\left(N\right)\propto \prod_{N'=N_{\min}}^{N-3} \frac{ Q\left(N' \to N'+3\right)}
{ Q\left(N'+3 \to N'\right)}\sim e^{-L \beta G\left(\mu,r\right)},
\end{equation}
where $N_{\min}=N-3 \min_{\alpha} \left( N_{\alpha}\right)$ is the minimal number of particles.
The LDF of the overall density, $r$, is
\begin{equation}
\label{eq:LDF}
G(\mu,r) =  -\mu r +\int_{r_0}^r dr' \mu \left( r' \right),
\end{equation}
where $r_0$ is an arbitrary parameter and
\begin{equation}
\label{eq:mu_def}
\mu(r) = \frac{1}{3\beta} \Bigl[ \log\Bigl( \int_0^1dx
\rho_{A}^\star\rho_B^\star\rho_C^\star \Bigr) -
3 \log\left( 1-r \right) \Bigr].
\end{equation}
The LDF, which is in fact proportional to the potential felt by the random walker, is plotted in \fref{fig:Fmu} for some point in parameter-space. In the large $L$ limit, the average value of $r$
is given by the global minimum of $G\left(\mu,r\right)$ which corresponds to
$\partial G/\partial r=\mu-\mu\left(r\right)=0$.

We now discuss briefly the role of $\mu(r)$ as the chemical potential of the conserving model.
This will enable us to compare the phase diagrams of the two models in the $(\beta,\mu)$-plane.
In the absence of a Hamiltonian which states the energy cost of adding or removing particles,
the chemical potential of a conserving nonequilibrium system can be measured by coupling it
to a microscopic gauge, customarily defined using Creutz method \cite{Creutz1983}.
Here, this is done by allowing triplets of $ABC$ particles to evaporate from the lattice into a {\it demon} with the slow rate $p$ defined in \eref{eq:slowp}.
If the demon contains triplets, it may depose them
back into the lattice at the same rate $p$.
Following the lines of the derivation above, the probability density of the number particles in the demon, $N_d\geq0$, can be shown to obey in the limit $L\to \infty$,
\begin{equation}
\label{eq:daemon}
\frac{P\left(N_{d}+3\right)}{P\left(N_d\right)} \simeq
\frac{\int_0^1dx\rho_{A}^\star\rho_B^\star\rho_C^\star}{\int_0^1 dx\left[\rho_0^\star\right]^3}= e^{3\beta\mu\left(r\right)},
\end{equation}
and hence $P\left(N_d\right)\propto \exp\left({\beta \mu \left(r\right) N_d}\right)$ \footnote{
For $\mu(r)<0$ we obtain $N_d\sim O(1)$ and $r=(N-N_d)/L$ fixed, as assumed in \eref{eq:daemon}. For $\mu(r)>0$, we need to consider $N_d\leq0$, by allowing the demon to store $000$ triplets instead of $ABC$ triplets.}.
The function $\mu\left(r\right)$ is therefore the chemical potential of the conserving model, as measured by the demon.
 In contrast to equilibrium, here $\mu(r)$ is derived from $W(\boldsymbol \zeta,N | \boldsymbol \zeta',N')$ and thus depends on the choice of nonconserving dynamics.

\begin{figure}
\noindent
\begin{centering}\includegraphics[scale=0.46]{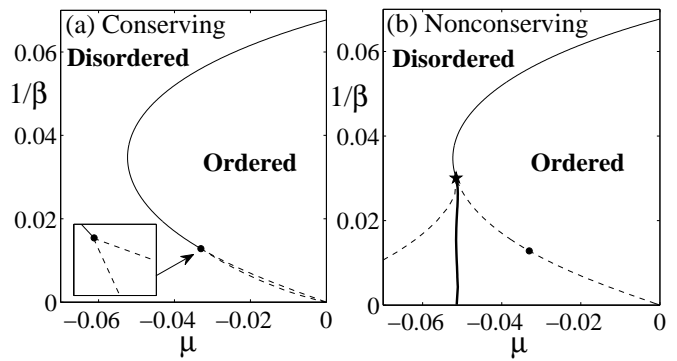}\par\end{centering}
\caption{ The $(1/\beta,\mu)$ phase diagrams of the conserving (a) and nonconserving (b) models for two equal densities with $\Delta=0.025$.
In the conserving model $\mu$ is computed from \eref{eq:mu_def}.
 The thick and thin solid lines
 represent the first and second order phase transitions, respectively. They join
 at the conserving ($\bullet$) and nonconserving ($\star$) tricritical points.
 The dashed lines denote the stability limits of the two phases.
 The inset in (a) depicts schematically the area near the conserving tricritical point.
The conserving tricritical point is irrelevant in (b),
 as it is located within the ordered phase of the nonconserving model.
\label{fig:mut}}
\end{figure}

We now proceed to compute the
 phase diagrams of the model under conserving dynamics and slow nonconserving
 dynamics.
 The conserving phase diagram ($p=0,r\leq1$) can be derived
from that of the original ABC model ($p=0,r=1$) using a mapping where the vacant sites are removed from each microstate of the conserving model. The master equation of the resulting system corresponds to that of the original ABC model
with $N$ sites. By observing that $q=\exp(\beta/L)=\exp(\beta r/N)\equiv \exp(\beta'/N) $ we conclude that the $N$-size
system has an effective bias of $\beta'=\beta r$. Similarly the average
densities of the $N$-size system can be shown to be given by $r_\alpha'=r_\alpha/r$ for
$\alpha=A,B,C$ and $r'_0=0$. The steady-state profile of the conserving model can thus be
expressed as
\begin{equation}
\label{eq:mapping}
\rho_\alpha^\star\left(x,\beta,r_{\alpha}, r\right)=r\rho_\alpha^\star\left(x,\beta r, r_{\alpha}/r,1\right),
\end{equation}
where $\rho^\star_\alpha$ in the r.h.s. has been derived for arbitrary $r_\alpha$ in \cite{Cohen2011a}.
\Eref{eq:mapping} maps the phase diagram of the original ABC model \cite{Clincy2003,Cohen2011a}
onto the conserving model ($r\leq1, p=0)$.
The resulting conserving phase diagram consists of a second order transition line at
\begin{equation}
\label{eq:bc}
\beta=2\pi\sqrt{3}/\sqrt{r^2-36\Delta^2},
\end{equation}
where $\Delta^2=\frac{1}{6}\sum_{\alpha=A,B,C}\left(r_\alpha-r/3\right)^2$ is a measure
for the deviation from equal densities. The transition becomes first order
for $\left(r_A^2+r_B^2+r_C^2\right)r > 2 \left( r_A^3+r_B^3+r_C^3\right)$.
 The phase diagram is shown in \fref{fig:mut}a for two equal densities,
\begin{equation}
\label{eq:twoequal}
r_A=r_B=r/3-\Delta, \qquad r_C=r/3+2\Delta.
\end{equation}
Since the LDF of $\rho_{\alpha}\left(x\right)$ in not known, one cannot compute the first
order transition line of the conserving model.
 It is possible, however, to draw the stability limits (dashed lines) in between which both phases are stable with respect to small perturbations.

\begin{figure}[t]
\noindent
\begin{centering}\includegraphics[scale=0.46]{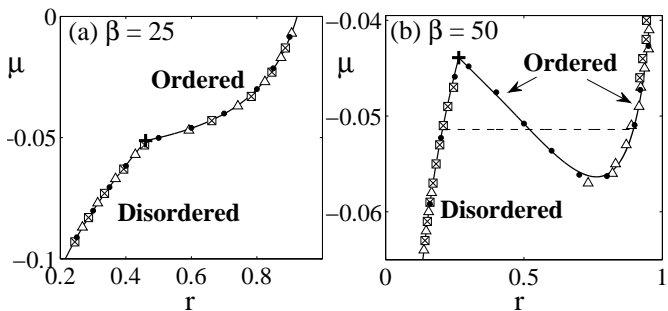}\par\end{centering}
\caption{ The $\mu(r)$ curve for two equal densities
with $\Delta=0.025$ and two values $\beta$. The hydrodynamic
solutions of the homogenous and ordered phases (solid lines) meet at the critical point ($+$).
They are shown in comparison with the results of the conserving simulation ($\bullet$)
and nonconserving simulation with homogenous ($\boxtimes$) and fully ordered ($\triangle$) initial states,
performed with $L=2400$ and $p=0.0001$. The dashed line in (b) denotes the nonconserving first order transition point.
\label{fig:mu_r}}
\end{figure}

The phase diagram of the nonconserving model can be derived by studying the
extrema of $G\left(\mu,r\right)$, given by the equation $\mu=\mu(r)$. The function $\mu(r)$, defined in
\eref{eq:mu_def}, is plotted for the two equal densities case in \fref{fig:mu_r}.
At low values of $\beta$ (\fref{fig:mu_r}a) there is a one-to-one correspondence
between $\mu$ and $r$. The conserving and nonconserving models therefore behave similarly
 and display a second order phase transition.
At high values of $\beta$ (\fref{fig:mu_r}b) we
observe a region of $\mu$ where  $G\left(\mu,r\right)$ has three extrema.
The intermediate density extremum has negative compressibility and
corresponds to a maximum of $G\left(\mu,r\right)$. It
is thus stable only in the conserving model, while the nonconserving
model undergoes a first order transition according to the Maxwell's construction (dashed line). The construction is justified by analyzing Eqs. (\ref{eq:PN}) and (\ref{eq:LDF}) at the equal area point where
 \begin{equation}
\frac{ P\left(r_{\text{ord}}^\star\right)}{P\left(r_{\text{hom}}^\star\right)}=\exp\Big[L\beta\int_{r_{\text{hom}}^\star}^{r_{\text{ord}}^\star}dr'\left(\mu-\mu(r')\right)\Big]=1.
 \end{equation}
 Here $r_{\text{ord}}^\star$ and $r_{\text{hom}}^\star$ denote the value of $r$
 at the minima of $G(\mu,r)$ that correspond to the ordered and homogenous phases, respectively.
 This method, employed for various values of $\beta$, yields the first order transition line (thick line in \fref{fig:mut}b).
The right and left dashed lines in \fref{fig:mut}b correspond to the stability limit of the
 homogenous and ordered phases, respectively. They define the coexistence region where
  we find three extrema of $G\left(\mu,r\right)$.

The picture emerging from the derivation above can be verified using Monte Carlo simulations, displayed in
\fref{fig:mu_r}.
In the conserving simulation, consisting of processes (\ref{eq:ABCdynamics}) and (\ref{eq:vacancyexchange}), $\mu(r)$ is evaluated by measuring the average number of ABC triplets in the lattice. The results show good agreement
with the hydrodynamic solution, thus confirming the validity of \eref{eq:meanfieldA}.
 In the nonconserving simulations, consisting of processes (\ref{eq:ABCdynamics})-(\ref{eq:addremoverates}), the average value of $r$ is measured.
The results deviate from the theoretical curve
only close to the first order transition shown in \fref{fig:mu_r}b.
There, we find different values of $r$ depending
  on whether the simulation was initiated in the fully phase separated or homogeneous states.
This hysteretic behaviour is an indication of a first order transition.
 The agreement with the theoretical results demonstrates the applicability of the large $L$ limit considered above to finite systems with slow nonconserving processes.

To conclude, we have studied the generalized ABC model with slow nonconserving dynamics.
This limit enables us to derive
 an exact expression for the LDF of the overall density, $r$, based on the knowledge of the
 conserving steady-state, $\rho_\alpha^\star(x,r)$, despite the fact that the LDF of $\rho_\alpha(x)$
 is not known.
 In addition, we define the chemical potential of the model, which unlike the equilibrium one, depends
 on the details of the nonconserving dynamics.
Based on this approach we compute the exact phase diagrams of the conserving and nonconserving models.
They consist of a second order transition line which turns into a first order line at
different tricritical points in each model. Such ensemble inequivalence
is typical of equilibrium models with long-range interactions.
This suggests that due to long-range correlations, which appear generically
in driven diffusive systems \cite{Spohn1983,Garrido1990,Dorfman1994,Schmittmann1995,Ortiz2004,Sadhu2011}, the
`grand-canonical' phase diagram derived following the approach presented above may
often differ from the corresponding `canonical' phase diagram.

The derivation above can be readily applied to other driven models that are
coupled slowly to an external bath.
The explicit expression of the LDF of the nonconserved parameter may, however,
depend on the steady state properties of the conserving model which are known analytically only in a handful of models.
We demonstrate this approach and the significance of the choice of the nonconserving process by considering the generalized ABC model with the usual grand-canonical dynamics, where the nonconserving process (\ref{eq:addremoverates}) is replaced by
\begin{equation}
A\overset{p e^{-\beta\mu}}{\underset{p}{\rightleftarrows}}0,
\qquad
B\overset{p e^{-\beta\mu}}{\underset{p}{\rightleftarrows}}0,
\qquad
C\overset{p e^{-\beta\mu}}{\underset{p}{\rightleftarrows}}0.
\end{equation}
Assuming slow nonconserving dynamics (\ref{eq:slowp}) and following the derivation above yields the
same LDF as in \eref{eq:LDF} but with $\mu(r)=[\log(r)-\log(1-r)]/\beta$.
As expected, different nonconserving dynamics lead to different definitions of the
chemical potential.
 Here, since $\mu(r)$ is single-valued for any value of $\beta$,
 the conserving and nonconserving model display the same phase diagram. It
consists of a second order transition line given by \eref{eq:bc} for $\Delta=0$.

It would be interesting to study the borderline case of $\gamma=2$ and investigate how the picture presented above changes when the `adiabatic' approximation (\ref{eq:adiabatic}) breaks down.
A similar limit has recently been studied in a boundary-driven diffusive model with nonconserving dynamics in the bulk,
for which an implicit expression for the LDF of the profile was derived \cite{Bodineau2010}.
In a different study, an approximate chemical potential has been defined for several driven models where
the nonconserving dynamics is not slow \cite{Pradhan2010,Pradhan2011,Pradhan2011a}.

\begin{acknowledgements}
We thank A. Bar, M. R. Evans, O. Hirschberg, A. Lederhendler, T. Sadhu and Y. Shokef
for helpful discussions. The support of the Israel Science
Foundation (ISF) is gratefully acknowledged.
\end{acknowledgements}

\bibliographystyle{apsrev4-1}

\bibliography{ABCModel}

%Merlin.mbs v4.21 2009-07-09.
\begin{thebibliography}{10}%
\makeatletter
\providecommand \@ifxundefined [1]{%
 \ifx #1\undefined \expandafter \@firstoftwo
 \else \expandafter \@secondoftwo
\fi
}%
\providecommand \@ifnum [1]{%
 \ifnum #1\expandafter \@firstoftwo
 \else \expandafter \@secondoftwo
\fi
}%
\providecommand \enquote [1]{``#1''}%
\providecommand \bibnamefont  [1]{#1}%
\providecommand \bibfnamefont [1]{#1}%
\providecommand \citenamefont [1]{#1}%
\providecommand\href[0]{\@sanitize\@href}%
\providecommand\@href[1]{\endgroup\@@startlink{#1}\endgroup\@@href}%
\providecommand\@@href[1]{#1\@@endlink}%
\providecommand \@sanitize [0]{\begingroup\catcode`\&12\catcode`\#12\relax}%
\@ifxundefined \pdfoutput {\@firstoftwo}{%
 \@ifnum{\z@=\pdfoutput}{\@firstoftwo}{\@secondoftwo}%
}{%
 \providecommand\@@startlink[1]{\leavevmode\special{html:<a href="#1">}}%
 \providecommand\@@endlink[0]{\special{html:</a>}}%
}{%
 \providecommand\@@startlink[1]{%
  \leavevmode
  \pdfstartlink
   attr{/Border[0 0 1 ]/H/I/C[0 1 1]}%
   user{/Subtype/Link/A<</Type/Action/S/URI/URI(#1)>>}%
  \relax
 }%
 \providecommand\@@endlink[0]{\pdfendlink}%
}%
\providecommand \url  [0]{\begingroup\@sanitize \@url }%
\providecommand \@url [1]{\endgroup\@href {#1}{\urlprefix}}%
\providecommand \urlprefix [0]{URL }%
\providecommand \Eprint[0]{\href }%
\@ifxundefined \urlstyle {%
  \providecommand \doi [1]{doi:\discretionary{}{}{}#1}%
}{%
  \providecommand \doi [0]{doi:\discretionary{}{}{}\begingroup
  \urlstyle{rm}\Url }%
}%
\providecommand \doibase [0]{http://dx.doi.org/}%
\providecommand \Doi[1]{\href{\doibase#1}}%
\providecommand \bibAnnote [3]{%
  \BibitemShut{#1}%
  \begin{quotation}\noindent
    \textsc{Key:}\ #2\\\textsc{Annotation:}\ #3%
  \end{quotation}%
}%
\providecommand \bibAnnoteFile [2]{%
  \IfFileExists{#2}{\bibAnnote {#1} {#2} {\input{#2}}}{}%
}%
\providecommand \typeout [0]{\immediate \write \m@ne }%
\providecommand \selectlanguage [0]{\@gobble}%
\providecommand \bibinfo [0]{\@secondoftwo}%
\providecommand \bibfield [0]{\@secondoftwo}%
\providecommand \translation [1]{[#1]}%
\providecommand \BibitemOpen[0]{}%
\providecommand \bibitemStop [0]{}%
\providecommand \bibitemNoStop [0]{.\EOS\space}%
\providecommand \EOS [0]{\spacefactor3000\relax}%
\providecommand \BibitemShut [1]{\csname bibitem#1\endcsname}%
%</preamble>
\bibitem{Mukamel2000}%
  \BibitemOpen
  \bibfield{author}{%
  \bibinfo {author} {\bibfnamefont{D.}~\bibnamefont{Mukamel}},\ }%
  in\ \emph{\bibinfo {booktitle} {Soft and Fragile Matter: Metastability and
  Flow}},\ \bibinfo {editor} {edited by\ \bibinfo {editor}
  {\bibfnamefont{M.~E.}\ \bibnamefont{Cates}}\ and\ \bibinfo {editor}
  {\bibfnamefont{M.~R.}\ \bibnamefont{Evans}}}\ (\bibinfo {publisher} {Bristol:
  Institute of Physics Publishing},\ \bibinfo {year} {2000})%
  \bibAnnoteFile{NoStop}{Mukamel2000}%
\bibitem{Schutz2001}%
  \BibitemOpen
  \bibfield{author}{%
  \bibinfo {author} {\bibfnamefont{G.~M.}\ \bibnamefont{Sch\"utz}},\ }%
  in\ \emph{\bibinfo {booktitle} {Phase Transitions and Critical Phenomena}},\
  Vol.~\bibinfo {volume} {19},\ \bibinfo {editor} {edited by\ \bibinfo {editor}
  {\bibfnamefont{C.}~\bibnamefont{Domb}}\ and\ \bibinfo {editor}
  {\bibfnamefont{J.~L.}\ \bibnamefont{Lebowitz}}}\ (\bibinfo {publisher}
  {Academic Press, London},\ \bibinfo {year} {2000})\ pp.\ \bibinfo {pages}
  {1--251}%
  \bibAnnoteFile{NoStop}{Schutz2001}%
\bibitem{Evans2005}%
  \BibitemOpen
  \bibfield{author}{%
  \bibinfo {author} {\bibfnamefont{M.~R.}\ \bibnamefont{Evans}}\ and\ \bibinfo
  {author} {\bibfnamefont{T.}~\bibnamefont{Hanney}},\ }%
  \bibfield{journal}{%
  \Doi{10.1088/0305-4470/38/19/R01}{\bibinfo {journal} {Journal of Physics A:
  Mathematical and General}}\ }%
  \textbf{\bibinfo {volume} {38}},\ \bibinfo {pages} {R195} (\bibinfo {year}
  {2005})%
  \bibAnnoteFile{NoStop}{Evans2005}%
\bibitem{Schadschneider2010}%
  \BibitemOpen
  \bibfield{author}{%
  \bibinfo {author} {\bibfnamefont{A.}~\bibnamefont{Schadschneider}}, \bibinfo
  {author} {\bibfnamefont{D.}~\bibnamefont{Chowdhury}},\ and\ \bibinfo {author}
  {\bibfnamefont{K.}~\bibnamefont{Nishinari}},\ }%
  \emph{\bibinfo {title} {Stochastic Transport in Complex Systems: From
  Molecules to Vehicles}}\ (\bibinfo {publisher} {Elsevier, New York},\
  \bibinfo {year} {2010})%
  \bibAnnoteFile{NoStop}{Schadschneider2010}%
\bibitem{Evans2002}%
  \BibitemOpen
  \bibfield{author}{%
  \bibinfo {author} {\bibfnamefont{M.~R.}\ \bibnamefont{Evans}}, \bibinfo
  {author} {\bibfnamefont{Y.}~\bibnamefont{Kafri}}, \bibinfo {author}
  {\bibfnamefont{E.}~\bibnamefont{Levine}},\ and\ \bibinfo {author}
  {\bibfnamefont{D.}~\bibnamefont{Mukamel}},\ }%
  \bibfield{journal}{%
  \Doi{10.1088/0305-4470/35/29/101}{\bibinfo {journal} {Journal of Physics A:
  Mathematical and General}}\ }%
  \textbf{\bibinfo {volume} {35}},\ \bibinfo {pages} {L433} (\bibinfo {year}
  {2002})%
  \bibAnnoteFile{NoStop}{Evans2002}%
\bibitem{Willmann2002}%
  \BibitemOpen
  \bibfield{author}{%
  \bibinfo {author} {\bibfnamefont{R.~D.}\ \bibnamefont{Willmann}}, \bibinfo
  {author} {\bibfnamefont{G.~M.}\ \bibnamefont{Sch\"utz}},\ and\ \bibinfo
  {author} {\bibfnamefont{D.}~\bibnamefont{Challet}},\ }%
  \bibfield{journal}{%
  \Doi{10.1016/S0378-4371(02)01217-7}{\bibinfo {journal} {Physica A}}\ }%
  \textbf{\bibinfo {volume} {316}},\ \bibinfo {pages} {430 } (\bibinfo {year}
  {2002})%
  \bibAnnoteFile{NoStop}{Willmann2002}%
\bibitem{Parmeggiani2003}%
  \BibitemOpen
  \bibfield{author}{%
  \bibinfo {author} {\bibfnamefont{A.}~\bibnamefont{Parmeggiani}}, \bibinfo
  {author} {\bibfnamefont{T.}~\bibnamefont{Franosch}},\ and\ \bibinfo {author}
  {\bibfnamefont{E.}~\bibnamefont{Frey}},\ }%
  \bibfield{journal}{%
  \Doi{10.1103/PhysRevLett.90.086601}{\bibinfo {journal} {Phys. Rev. Lett.}}\
  }%
  \textbf{\bibinfo {volume} {90}},\ \bibinfo {pages} {086601} (\bibinfo {year}
  {2003})%
  \bibAnnoteFile{NoStop}{Parmeggiani2003}%
\bibitem{Evans2003}%
  \BibitemOpen
  \bibfield{author}{%
  \bibinfo {author} {\bibfnamefont{M.~R.}\ \bibnamefont{Evans}}, \bibinfo
  {author} {\bibfnamefont{R.}~\bibnamefont{Juh\'asz}},\ and\ \bibinfo {author}
  {\bibfnamefont{L.}~\bibnamefont{Santen}},\ }%
  \bibfield{journal}{%
  \Doi{10.1103/PhysRevE.68.026117}{\bibinfo {journal} {Phys. Rev. E}}\ }%
  \textbf{\bibinfo {volume} {68}},\ \bibinfo {pages} {026117} (\bibinfo {year}
  {2003})%
  \bibAnnoteFile{NoStop}{Evans2003}%
\bibitem{Popkov2003}%
  \BibitemOpen
  \bibfield{author}{%
  \bibinfo {author} {\bibfnamefont{V.}~\bibnamefont{Popkov}}, \bibinfo {author}
  {\bibfnamefont{A.}~\bibnamefont{R\'akos}}, \bibinfo {author}
  {\bibfnamefont{R.~D.}\ \bibnamefont{Willmann}}, \bibinfo {author}
  {\bibfnamefont{A.~B.}\ \bibnamefont{Kolomeisky}},\ and\ \bibinfo {author}
  {\bibfnamefont{G.~M.}\ \bibnamefont{Sch\"utz}},\ }%
  \bibfield{journal}{%
  \Doi{10.1103/PhysRevE.67.066117}{\bibinfo {journal} {Phys. Rev. E}}\ }%
  \textbf{\bibinfo {volume} {67}},\ \bibinfo {pages} {066117} (\bibinfo {year}
  {2003})%
  \bibAnnoteFile{NoStop}{Popkov2003}%
\bibitem{Levine2004}%
  \BibitemOpen
  \bibfield{author}{%
  \bibinfo {author} {\bibfnamefont{E.}~\bibnamefont{Levine}}\ and\ \bibinfo
  {author} {\bibfnamefont{R.~D.}\ \bibnamefont{Willmann}},\ }%
  \bibfield{journal}{%
  \Doi{doi:10.1088/0305-4470/37/10/002}{\bibinfo {journal} {Journal of Physics
  A: Mathematical and General}}\ }%
  \textbf{\bibinfo {volume} {37}},\ \bibinfo {pages} {3333} (\bibinfo {year}
  {2004})%
  \bibAnnoteFile{NoStop}{Levine2004}%
\bibitem{Sasa2006}%
  \BibitemOpen
  \bibfield{author}{%
  \bibinfo {author} {\bibfnamefont{S.}~\bibnamefont{Sasa}}\ and\ \bibinfo
  {author} {\bibfnamefont{H.}~\bibnamefont{Tasaki}},\ }%
  \bibfield{journal}{%
  \Doi{10.1007/s10955-005-9021-7}{\bibinfo {journal} {J. Stat. Phys.}}\ }%
  \textbf{\bibinfo {volume} {125}},\ \bibinfo {pages} {125} (\bibinfo {year}
  {2006})%
  \bibAnnoteFile{NoStop}{Sasa2006}%
\bibitem{Bodineau2010}%
  \BibitemOpen
  \bibfield{author}{%
  \bibinfo {author} {\bibfnamefont{T.}~\bibnamefont{Bodineau}}\ and\ \bibinfo
  {author} {\bibfnamefont{M.}~\bibnamefont{Lagouge}},\ }%
  \bibfield{journal}{%
  \Doi{10.1007/s10955-010-9934-7}{\bibinfo {journal} {J. Stat. Phys.}}\ }%
  \textbf{\bibinfo {volume} {139}},\ \bibinfo {pages} {201} (\bibinfo {year}
  {2010})%
  \bibAnnoteFile{NoStop}{Bodineau2010}%
\bibitem{Evans1998}%
  \BibitemOpen
  \bibfield{author}{%
  \bibinfo {author} {\bibfnamefont{M.~R.}\ \bibnamefont{Evans}}, \bibinfo
  {author} {\bibfnamefont{Y.}~\bibnamefont{Kafri}}, \bibinfo {author}
  {\bibfnamefont{H.~M.}\ \bibnamefont{Koduvely}},\ and\ \bibinfo {author}
  {\bibfnamefont{D.}~\bibnamefont{Mukamel}},\ }%
  \bibfield{journal}{%
  \Doi{10.1103/PhysRevLett.80.425}{\bibinfo {journal} {Phys. Rev. Lett.}}\ }%
  \textbf{\bibinfo {volume} {80}},\ \bibinfo {pages} {425} (\bibinfo {year}
  {1998})%
  \bibAnnoteFile{NoStop}{Evans1998}%
\bibitem{Evans1998b}%
  \BibitemOpen
  \bibfield{author}{%
  \bibinfo {author} {\bibfnamefont{M.~R.}\ \bibnamefont{Evans}}, \bibinfo
  {author} {\bibfnamefont{Y.}~\bibnamefont{Kafri}}, \bibinfo {author}
  {\bibfnamefont{H.~M.}\ \bibnamefont{Koduvely}},\ and\ \bibinfo {author}
  {\bibfnamefont{D.}~\bibnamefont{Mukamel}},\ }%
  \bibfield{journal}{%
  \Doi{10.1103/PhysRevE.58.2764}{\bibinfo {journal} {Phys. Rev. E}}\ }%
  \textbf{\bibinfo {volume} {58}},\ \bibinfo {pages} {2764} (\bibinfo {year}
  {1998})%
  \bibAnnoteFile{NoStop}{Evans1998b}%
\bibitem{Clincy2003}%
  \BibitemOpen
  \bibfield{author}{%
  \bibinfo {author} {\bibfnamefont{M.}~\bibnamefont{Clincy}}, \bibinfo {author}
  {\bibfnamefont{B.}~\bibnamefont{Derrida}},\ and\ \bibinfo {author}
  {\bibfnamefont{M.~R.}\ \bibnamefont{Evans}},\ }%
  \bibfield{journal}{%
  \Doi{10.1103/PhysRevE.67.066115}{\bibinfo {journal} {Phys. Rev. E}}\ }%
  \textbf{\bibinfo {volume} {67}},\ \bibinfo {pages} {066115} (\bibinfo {year}
  {2003})%
  \bibAnnoteFile{NoStop}{Clincy2003}%
\bibitem{Cohen2011a}%
  \BibitemOpen
  \bibfield{author}{%
  \bibinfo {author} {\bibfnamefont{O.}~\bibnamefont{Cohen}}\ and\ \bibinfo
  {author} {\bibfnamefont{D.}~\bibnamefont{Mukamel}},\ }%
  \bibfield{journal}{%
  \Doi{doi:10.1088/1742-5468/2010/11/P11016}{\bibinfo {journal} {J. Phys. A}}\
  }%
  \textbf{\bibinfo {volume} {44}},\ \bibinfo {pages} {415004} (\bibinfo {year}
  {2011})%
  \bibAnnoteFile{NoStop}{Cohen2011a}%
\bibitem{Ayyer2009}%
  \BibitemOpen
  \bibfield{author}{%
  \bibinfo {author} {\bibfnamefont{A.}~\bibnamefont{Ayyer}}, \bibinfo {author}
  {\bibfnamefont{E.~A.}\ \bibnamefont{Carlen}}, \bibinfo {author}
  {\bibfnamefont{J.~L.}\ \bibnamefont{Lebowitz}}, \bibinfo {author}
  {\bibfnamefont{P.~K.}\ \bibnamefont{Mohanty}}, \bibinfo {author}
  {\bibfnamefont{D.}~\bibnamefont{Mukamel}},\ and\ \bibinfo {author}
  {\bibfnamefont{E.~R.}\ \bibnamefont{Speer}},\ }%
  \bibfield{journal}{%
  \Doi{10.1007/s10955-009-9834-x}{\bibinfo {journal} {J. Stat. Phys.}}\ }%
  \textbf{\bibinfo {volume} {137}},\ \bibinfo {pages} {1166} (\bibinfo {year}
  {2009})%
  \bibAnnoteFile{NoStop}{Ayyer2009}%
\bibitem{Barton2011}%
  \BibitemOpen
  \bibfield{author}{%
  \bibinfo {author} {\bibfnamefont{J.}~\bibnamefont{Barton}}, \bibinfo {author}
  {\bibfnamefont{J.~L.}\ \bibnamefont{Lebowitz}},\ and\ \bibinfo {author}
  {\bibfnamefont{E.~R.}\ \bibnamefont{Speer}},\ }%
  \bibfield{journal}{%
  \bibinfo {journal} {ArXiv e-prints}}%
   (\bibinfo {year} {2011}),\
  \Eprint{http://arxiv.org/abs/1106.1942}{arXiv:1106.1942}%
  \bibAnnoteFile{NoStop}{Barton2011}%
\bibitem{Barton2010}%
  \BibitemOpen
  \bibfield{author}{%
  \bibinfo {author} {\bibfnamefont{J.}~\bibnamefont{Barton}}, \bibinfo {author}
  {\bibfnamefont{J.~L.}\ \bibnamefont{Lebowitz}},\ and\ \bibinfo {author}
  {\bibfnamefont{E.~R.}\ \bibnamefont{Speer}},\ }%
  \bibfield{journal}{%
  \Doi{doi:10.1088/1751-8113/44/6/065005}{\bibinfo {journal} {J. Phys. A}}\ }%
  \textbf{\bibinfo {volume} {44}},\ \bibinfo {pages} {065005} (\bibinfo {year}
  {2011})%
  \bibAnnoteFile{NoStop}{Barton2010}%
\bibitem{Lederhendler2010a}%
  \BibitemOpen
  \bibfield{author}{%
  \bibinfo {author} {\bibfnamefont{A.}~\bibnamefont{Lederhendler}}\ and\
  \bibinfo {author} {\bibfnamefont{D.}~\bibnamefont{Mukamel}},\ }%
  \bibfield{journal}{%
  \Doi{10.1103/PhysRevLett.105.150602}{\bibinfo {journal} {Phys. Rev. Lett.}}\
  }%
  \textbf{\bibinfo {volume} {105}},\ \bibinfo {pages} {150602} (\bibinfo {year}
  {2010})%
  \bibAnnoteFile{NoStop}{Lederhendler2010a}%
\bibitem{Lederhendler2010b}%
  \BibitemOpen
  \bibfield{author}{%
  \bibinfo {author} {\bibfnamefont{A.}~\bibnamefont{Lederhendler}}, \bibinfo
  {author} {\bibfnamefont{O.}~\bibnamefont{Cohen}},\ and\ \bibinfo {author}
  {\bibfnamefont{D.}~\bibnamefont{Mukamel}},\ }%
  \bibfield{journal}{%
  \Doi{doi:10.1088/1742-5468/2010/11/P11016}{\bibinfo {journal} {J. Stat. Mech:
  Theory Exp.}}\ }%
  \textbf{\bibinfo {volume} {2010}},\ \bibinfo {pages} {P11016} (\bibinfo
  {year} {2010})%
  \bibAnnoteFile{NoStop}{Lederhendler2010b}%
\bibitem{Dauxois2009}%
  \BibitemOpen
  \emph{\bibinfo {title} {Long-Range Interacting Systems (Les Houches Summer
  School 2008)}},\ edited by\ \bibinfo {editor}
  {\bibfnamefont{T.}~\bibnamefont{Dauxois}}, \bibinfo {editor}
  {\bibfnamefont{S.}~\bibnamefont{Ruffo}},\ and\ \bibinfo {editor}
  {\bibfnamefont{L.~F.}\ \bibnamefont{Cugliandolo}}\ (\bibinfo {publisher}
  {Oxford: Oxford University Press, New York},\ \bibinfo {year} {2009})%
  \bibAnnoteFile{NoStop}{Dauxois2009}%
\bibitem{Dauxois2010}%
  \BibitemOpen
  \emph{\bibinfo {title} {Topical issue: Long-Range Interacting Systems}},\
  edited by\ \bibinfo {editor} {\bibfnamefont{T.}~\bibnamefont{Dauxois}}\ and\
  \bibinfo {editor} {\bibfnamefont{S.}~\bibnamefont{Ruffo}}\ (\bibinfo
  {publisher} {J. Stat. Mech: Theory Exp.},\ \bibinfo {year} {2010})%
  \bibAnnoteFile{NoStop}{Dauxois2010}%
\bibitem{Creutz1983}%
  \BibitemOpen
  \bibfield{author}{%
  \bibinfo {author} {\bibfnamefont{M.}~\bibnamefont{Creutz}},\ }%
  \bibfield{journal}{%
  \Doi{10.1103/PhysRevLett.50.1411}{\bibinfo {journal} {Phys. Rev. Lett.}}\ }%
  \textbf{\bibinfo {volume} {50}},\ \bibinfo {pages} {1411} (\bibinfo {year}
  {1983})%
  \bibAnnoteFile{NoStop}{Creutz1983}%
\bibitem{Note1}%
  \BibitemOpen
  \bibinfo {note} {For $\mu (r)<0$ we obtain $N_d\sim O(1)$ and $r=(N-N_d)/L$
  fixed, as assumed in Eq.\ (\ref {eq:daemon}). For $\mu (r)>0$, we need to
  consider $N_d\leq 0$, by allowing the demon to store $000$ triplets instead
  of $ABC$ triplets.}%
  \bibAnnoteFile{Stop}{Note1}%
\bibitem{Spohn1983}%
  \BibitemOpen
  \bibfield{author}{%
  \bibinfo {author} {\bibfnamefont{H.}~\bibnamefont{Spohn}},\ }%
  \bibfield{journal}{%
  \Doi{doi:10.1088/0305-4470/16/18/029}{\bibinfo {journal} {J. Phys. A}}\ }%
  \textbf{\bibinfo {volume} {16}},\ \bibinfo {pages} {4275} (\bibinfo {year}
  {1983})%
  \bibAnnoteFile{NoStop}{Spohn1983}%
\bibitem{Garrido1990}%
  \BibitemOpen
  \bibfield{author}{%
  \bibinfo {author} {\bibfnamefont{P.~L.}\ \bibnamefont{Garrido}}, \bibinfo
  {author} {\bibfnamefont{J.~L.}\ \bibnamefont{Lebowitz}}, \bibinfo {author}
  {\bibfnamefont{C.}~\bibnamefont{Maes}},\ and\ \bibinfo {author}
  {\bibfnamefont{H.}~\bibnamefont{Spohn}},\ }%
  \bibfield{journal}{%
  \Doi{10.1103/PhysRevA.42.1954}{\bibinfo {journal} {Phys. Rev. A}}\ }%
  \textbf{\bibinfo {volume} {42}},\ \bibinfo {pages} {1954} (\bibinfo {year}
  {1990})%
  \bibAnnoteFile{NoStop}{Garrido1990}%
\bibitem{Dorfman1994}%
  \BibitemOpen
  \bibfield{author}{%
  \bibinfo {author} {\bibfnamefont{J.~R.}\ \bibnamefont{Dorfman}}, \bibinfo
  {author} {\bibfnamefont{T.~R.}\ \bibnamefont{Kirkpatrick}},\ and\ \bibinfo
  {author} {\bibfnamefont{J.~V.}\ \bibnamefont{Sengers}},\ }%
  \bibfield{journal}{%
  \Doi{10.1146/annurev.pc.45.100194.001241}{\bibinfo {journal} {Annu. Rev.
  Phys. Chem.}}\ }%
  \textbf{\bibinfo {volume} {45}},\ \bibinfo {pages} {213} (\bibinfo {year}
  {1994})%
  \bibAnnoteFile{NoStop}{Dorfman1994}%
\bibitem{Schmittmann1995}%
  \BibitemOpen
  \bibfield{author}{%
  \bibinfo {author} {\bibfnamefont{B.}~\bibnamefont{Schmittmann}}\ and\
  \bibinfo {author} {\bibfnamefont{R.~K.~P.}\ \bibnamefont{Zia}},\ }%
  in\ \emph{\bibinfo {booktitle} {Statistical Mechanics of Driven Diffusive
  Systems}},\ Vol.~\bibinfo {volume} {17},\ \bibinfo {editor} {edited by\
  \bibinfo {editor} {\bibfnamefont{C.}~\bibnamefont{Domb}}\ and\ \bibinfo
  {editor} {\bibfnamefont{J.~L.}\ \bibnamefont{Lebowitz}}}\ (\bibinfo
  {publisher} {Academic Press, London},\ \bibinfo {year} {1995})%
  \bibAnnoteFile{NoStop}{Schmittmann1995}%
\bibitem{Ortiz2004}%
  \BibitemOpen
  \bibfield{author}{%
  \bibinfo {author} {\bibfnamefont{J.~M.}\ \bibnamefont{Ortiz~de Z{\'a}rate}}\
  and\ \bibinfo {author} {\bibfnamefont{J.~V.}\ \bibnamefont{Sengers}},\ }%
  \bibfield{journal}{%
  \Doi{10.1023/B:JOSS.0000028062.57459.52}{\bibinfo {journal} {J. Stat.
  Phys.}}\ }%
  \textbf{\bibinfo {volume} {115}},\ \bibinfo {pages} {1341} (\bibinfo {year}
  {2004})%
  \bibAnnoteFile{NoStop}{Ortiz2004}%
\bibitem{Sadhu2011}%
  \BibitemOpen
  \bibfield{author}{%
  \bibinfo {author} {\bibfnamefont{T.}~\bibnamefont{Sadhu}}, \bibinfo {author}
  {\bibfnamefont{S.~N.}\ \bibnamefont{Majumdar}},\ and\ \bibinfo {author}
  {\bibfnamefont{D.}~\bibnamefont{Mukamel}},\ }%
  \bibfield{journal}{%
  \bibinfo {journal} {ArXiv e-prints}}%
   (\bibinfo {year} {2011}),\
  \Eprint{http://arxiv.org/abs/1106.1838}{arXiv:1106.1838}%
  \bibAnnoteFile{NoStop}{Sadhu2011}%
\bibitem{Pradhan2010}%
  \BibitemOpen
  \bibfield{author}{%
  \bibinfo {author} {\bibfnamefont{P.}~\bibnamefont{Pradhan}}, \bibinfo
  {author} {\bibfnamefont{C.~P.}\ \bibnamefont{Amann}},\ and\ \bibinfo {author}
  {\bibfnamefont{U.}~\bibnamefont{Seifert}},\ }%
  \bibfield{journal}{%
  \Doi{10.1103/PhysRevLett.105.150601}{\bibinfo {journal} {Phys. Rev. Lett.}}\
  }%
  \textbf{\bibinfo {volume} {105}},\ \bibinfo {pages} {150601} (\bibinfo {year}
  {2010})%
  \bibAnnoteFile{NoStop}{Pradhan2010}%
\bibitem{Pradhan2011}%
  \BibitemOpen
  \bibfield{author}{%
  \bibinfo {author} {\bibfnamefont{P.}~\bibnamefont{Pradhan}}, \bibinfo
  {author} {\bibfnamefont{R.}~\bibnamefont{Ramsperger}},\ and\ \bibinfo
  {author} {\bibfnamefont{U.}~\bibnamefont{Seifert}},\ }%
  \bibfield{journal}{%
  \Doi{10.1103/PhysRevE.84.041104}{\bibinfo {journal} {Phys. Rev. E}}\ }%
  \textbf{\bibinfo {volume} {84}},\ \bibinfo {pages} {041104} (\bibinfo {year}
  {2011})%
  \bibAnnoteFile{NoStop}{Pradhan2011}%
\bibitem{Pradhan2011a}%
  \BibitemOpen
  \bibfield{author}{%
  \bibinfo {author} {\bibfnamefont{P.}~\bibnamefont{Pradhan}}\ and\ \bibinfo
  {author} {\bibfnamefont{U.}~\bibnamefont{Seifert}},\ }%
  \bibfield{journal}{%
  \Doi{10.1103/PhysRevE.84.051130}{\bibinfo {journal} {Phys. Rev. E}}\ }%
  \textbf{\bibinfo {volume} {84}},\ \bibinfo {pages} {051130} (\bibinfo {year}
  {2011})%
  \bibAnnoteFile{NoStop}{Pradhan2011a}%
\end{thebibliography}%

\end{document}